\documentclass[12pt]{iopart}
\usepackage{amssymb}
\usepackage{amsfonts}
\usepackage[english]{babel}
\usepackage{euscript}
\usepackage{bbm}

\begin{document}

\title[Deformations of Lie algebras, exotic cohomology and integrable  PDEs. II]{%
Deformations of infinite-dimensional Lie algebras, exotic cohomology and integrable nonlinear partial
differential equations. II}
\author{Oleg I. Morozov}

\begin{abstract}
We consider the four-dimensional reduced quasi-classical self-dual Yang--Mills equation
and show that non-triviality of the second exotic cohomology group of its
symmetry algebra implies existence of a two-component integrable generalization of this equation.
The sequence of natural extensions of this symmetry algebra generate an integrable hierarchy of multi-dimensional
nonlinear {\sc pde}s. We write out the first three elements of this hierarchy.
\end{abstract}

\address{Faculty of Applied
  Mathematics, AGH University of Science and Technology,
  \\
  Al. Mickiewicza 30,
  Cracow 30-059, Poland
\\
morozov{\symbol{64}}agh.edu.pl}



\maketitle
\section{Introduction}

The purpose of this paper is to provide new examples of application of the method of
\cite{Morozov2017,Morozov2018}  to the problem of finding a Lax representation for a {\sc pde}.
Lax representations, also known as zero-curvature representations, Wahlquist--Estabrook prolongation structures,
inverse scattering transformations, or differential coverings, are of great importance in the theory of
integrable systems, \cite{Mikhailov,Zakharov1991,LNM810}. They are a key feature of integrable {\sc pde}s and a
starting setting for a number of techniques of studying them such as B\"acklund transformations, Darboux
transformations, recursion operators, nonlocal symmetries, and nonlocal conservation laws. Although these
structures are of crucial  significance in the theory of integrable systems, up to now the problem of finding
conditions for a {\sc pde} to admit a Lax representation is unsolved. In \cite{Morozov2017,Morozov2018} we
propose a method for solving this problem in internal terms of the {\sc pde} under the study. The method uses
the second exotic cohomology group of the symmetry algebra of the {\sc pde} and allows one to get rid of apriori
assumptions about the form of the covering equations of a Lax representatation.

In the overwhelming majority of known examples, the symmetry algebras of in\-te\-gra\-ble multi-dimensional {\sc pde}s
are in\-fi\-ni\-te-di\-men\-si\-o\-nal. Computation of exotic cohomology groups for
infinite-dimensional Lie algebras is a complicated problem, while in some cases its so\-lu\-ti\-on is possible
owing to the specific graded structures of the algebras, \cite{Morozov2018}.  In this paper we show that
existence of a Lax representation can be derived from non-triviality of the se\-cond exotic cohomology group of
a finite-dimensional subalgebra of the symmetry al\-geb\-ra of the {\sc pde}.

Another important problem in the theory of integrable equations is concerned with the
integrable hierarchies associated with the {\sc pde} under the study. In \cite{Morozov2018} we
de\-mon\-stra\-te that
in some cases the graded structure of the symmetry algebra allows one to find such hierarchies.

In this paper we consider equation
\begin{equation}
u_{yz} = u_{tx}+u_y\,u_{xx}-u_x\,u_{xy},
\label{FKh4}
\end{equation}
which arise as a reduction of the quasi-classical self-dual Yang--Mills equation,
\cite{ManakovSantini2006,ManakovSantini2014,BKMV2015,BogdanovPavlov2016},
\begin{equation}
u_{yz} = u_{tx}+u_x\,u_{zs}-u_z\,u_{xs}.
\label{MASh5}
\end{equation}
Equation (\ref{FKh4}) has a number of remarkable properties.  In \cite{FerapontovKhusnutdinova2004}
the Lax representation
\begin{equation}
\left\{
\begin{array}{lcl}
q_t &=& \lambda\,q_y-u_y\,q_x,
\\
q_z &=& (\lambda-u_x)\,q_x
\end{array}
\right.
\label{FKH4_covering_lambda}
\end{equation}
with the nonremovable parameter $\lambda$ for equation (\ref{FKh4}) was presented. In \cite{KruglikovMorozov2016} a B\"acklund transformation
between (\ref{FKh4}) and another reduction
\[
v_{ty} = v_z\,v_{xy} - v_y\,v_{xz}
\]
of equation (\ref{MASh5}) was found.
The  symmetry algebra  of equation (\ref{FKh4}), as an abstract Lie algebra, is  the semi-direct sum
$\mathfrak{s}_2 = \mathfrak{s}_\diamond \ltimes \mathfrak{s}_{2, \infty}$  of the finite-dimensional
sub\-al\-geb\-ra $\mathfrak{s}_\diamond$ and the infinite-dimensional ideal
$\mathfrak{s}_{2, \infty} = \mathbb{R}_2[h] \otimes \mathfrak{q}$,
which is the tensor product of the (associative commutative unital) algebra
$\mathbb{R}_2[h] = \mathbb{R}[h]/ \langle  h^3 = 0 \rangle$ of truncated po\-ly\-no\-mi\-als of degree less than
3 and the Lie algebra $\mathfrak{q}$ of the vector fields of the form $A(t,z)\,\partial_z$ on $\mathbb{R}^2$.
The second exotic cohomology group of the finite-dimensional Lie algebra $\mathfrak{s}_\diamond$ is
one-dimensional.
The generating cocycle of this cohomology group defines a one-dimensional extension
$\tilde{\mathfrak{s}}_\diamond$ of $\mathfrak{s}_\diamond$. This extension provides a Lax representation
(\ref{extension_2_of_FKH4_covering}) for a two-component generalization (\ref{FKh4_extended}) of equation
(\ref{FKh4}), while (\ref{FKH4_covering_lambda}) and (\ref{FKh4}) are their evi\-dent reductions.

The symmetry algebra $\mathfrak{s}_2$ admits a series of natural extensions
$\mathfrak{s}_m
= \mathfrak{s}_\diamond \ltimes \mathfrak{s}_{m,\infty}
= \mathfrak{s}_\diamond \ltimes (\mathbb{R}_m[h] \otimes \mathfrak{q})$, $m\ge 3$,
which preserve the finite-dimensional part $\mathfrak{s}_\diamond $ as well as its extension
$\tilde{\mathfrak{s}}_\diamond$.The Maurer--Cartan of the series of Lie algebras
$\tilde{\mathfrak{s}}_\diamond \ltimes \mathfrak{s}_{m,\infty}$
provide a series of systems of {\sc pde}s with their Lax representations.  We write out the first three
elements of this hierarchy of integrable equations.

\section{Preliminaries}\label{Preliminaries_section}

All considerations in this paper are local. All functions are assumed to be real-analytic.

\subsection{Symmetries and differential coverings of {\sc pde}s}

The relevant geometric formulation of Lax representations is based on the concept of dif\-fe\-ren\-ti\-al
covering of a {\sc pde} \cite{KrasilshchikVinogradov1984,KrasilshchikVinogradov1989}. In this subsection we
closely follow \cite{KrasilshchikVinogradov1989,VinogradovKrasilshchik1997} to pre\-sent the basic notions of
the theory of differential coverings.

Let $\pi \colon \mathbb{R}^n \times \mathbb{R}^m \rightarrow \mathbb{R}^n$,
$\pi \colon (x^1, \dots, x^n, u^1, \dots, u^m)$    $\mapsto (x^1, \dots, x^n)$ be a trivial bundle, and
$J^\infty(\pi)$ be the bundle of its jets of the infinite order. The local coordinates on $J^\infty(\pi)$ are
$(x^i,u^\alpha,u^\alpha_I)$, where $I=(i_1, \dots, i_n)$ are multi-indices, and for every local section
$f \colon \mathbb{R}^n \rightarrow \mathbb{R}^n \times \mathbb{R}^m$ of $\pi$ the corresponding infinite jet
$j_\infty(f)$ is a section $j_\infty(f) \colon \mathbb{R}^n \rightarrow J^\infty(\pi)$ such that
$u^\alpha_I(j_\infty(f))
=\displaystyle{\frac{\partial^{\#I} f^\alpha}{\partial x^I}}
=\displaystyle{\frac{\partial^{i_1+\dots+i_n} f^\alpha}{(\partial x^1)^{i_1}\dots (\partial x^n)^{i_n}}}$.
We put $u^\alpha = u^\alpha_{(0,\dots,0)}$. Also, in the case of $m=1$ and, e.g., $n=4$ we denote
$x^1 = t$, $x^2= x$, $x^3= y$, $x^4=z$, and $u^1_{(i,j,k,l)}=u_{{t \dots t}{x \dots x}{y \dots y}{z \dots z}}$
with $i$  times $t$, $j$  times $x$, $k$  times $y$, and $l$  times $z$.

\vskip 3 pt

The vector fields
\[
D_{x^k} = \frac{\partial}{\partial x^k} + \sum \limits_{\# I \ge 0} \sum \limits_{\alpha = 1}^m
u^\alpha_{I+1_{k}}\,\frac{\partial}{\partial u^\alpha_I},
\qquad k \in \{1,\dots,n\},
\]
with $I+1_k =(i_1,\dots, i_k,\dots, i_n)+1_k = (i_1,\dots, i_k+1,\dots, i_n)$  are referred to as
{\it total derivatives}. They commute everywhere on $J^\infty(\pi)$:  $[D_{x^i}, D_{x^j}] = 0$.

\vskip 10 pt

A system of {\sc pde}s $F_r(x^i,u^\alpha_I) = 0$, $\# I \le s$, $r \in \{1,\dots, \sigma\}$, of the order
$s \ge 1$ with $\sigma \ge 1$ defines the submanifold
$\EuScript{E} = \{(x^i,u^\alpha_I) \in J^\infty(\pi) \,\,\vert\,\, D_K(F_r(x^i,u^\alpha_I)) = 0, \,\, \# K \ge 0\}$
in $J^\infty(\pi)$.

\vskip 5 pt

The \emph{evolutionary derivation} associated to an arbitrary smooth function
$\varphi \colon J^\infty(\pi) \to \mathbb{R}^m $ is the vector field
\begin{equation}
  \mathbf{E}_{\varphi} = \sum \limits_{\# I \ge 0} \sum \limits_{\alpha = 1}^m
  D_I(\varphi^\alpha)\,\frac{\partial}{\partial u^\alpha_I}
  \label{evolution_differentiation}
\end{equation}
with $D_I=D_{(i_1,\dots\,i_n)} =D^{i_1}_{x^1} \circ \dots \circ
D^{i_n}_{x^n}$.

A function $\varphi \colon \EuScript{E} \to \mathbb{R}^m$ is called a
\emph{\textup{(}generator of an infinitesimal\textup{)} symmetry} of $\EuScript{E}$ when
$\mathbf{E}_{\varphi}(F_r) = 0$ on $\EuScript{E}$. The symmetry $\varphi$ is a
solution to the \emph{defining system}
\begin{equation}
  \ell_{\EuScript{E}}(\varphi) = 0,
  \label{defining_eqns}
\end{equation}
where $\ell_{\EuScript{E}} = \left.{\ell_F}\right|_{\EuScript{E}}$ with the matrix
differential operator
\begin{equation*}
\ell_F = \left(\sum \limits_{\# I \ge 0}\frac{\partial F_a}{\partial
    u^\alpha_I}\,D_I\right).
\end{equation*}
Solutions to (\ref{defining_eqns}) constitute the Lie algebra $Sym(\EuScript{E})$ with respect
to the \emph{Jacobi bracket} $[\varphi, \psi] = \mathbf{E}_{\varphi}(\psi)-\mathbf{E}_{\psi}(\varphi)$.
The subalgebra of \emph{contact symmetries} of $\EuScript{E}$ is
$Sym_1(\EuScript{E}) =Sym(\EuScript{E}) \cap C^\infty(J^1(\pi),\mathbb{R}^m)$.
In its turn this subalgebra contains the subalgebra $Sym_0(\EuScript{E})$ of point symmetries, whose generators
have the form $\varphi=(\varphi^1, \dots, \varphi^m)$ with
\[
\varphi^\alpha = \eta^\alpha - \sum \limits_{j=1}^n \xi^j \,u^\alpha_{x^j},
\]
where $\eta^\alpha$, $\xi^i$ are functions of $J^0(\pi)$. Such generators  are in one-to-one correspondence
with the vector fields
\begin{equation}
\hat{\varphi} = \sum \limits_{j=1}^n \xi^j \frac{\partial}{\partial x^j} + \sum \limits_{\alpha=1}^m \eta^\alpha \frac{\partial}{\partial u^\alpha}
\label{hat_rule}
\end{equation}
 on $J^0(\pi)$.

\vskip  5 pt
Denote $\mathcal{W} = \mathbb{R}^\infty$ with  coordinates $w^a$, $a \in  \mathbb{N} \cup \{0\}$. Locally,
an (infinite-di\-men\-si\-o\-nal)  {\it differential covering} over $\EuScript{E}$ is a trivial bundle
$\tau \colon J^\infty(\pi) \times \mathcal{W} \rightarrow J^\infty(\pi)$
equipped with the {\it extended total derivatives}
\begin{equation}
\tilde{D}_{x^k} = D_{x^k} + \sum \limits_{a=0}^\infty
T^a_k(x^i,u^\alpha_I,w^b)\,\frac{\partial}{\partial w^a}
\label{extended_derivatives}
\end{equation}
such that $[\tilde{D}_{x^i}, \tilde{D}_{x^j}]=0$ for all $i \not = j$ whenever
$(x^i,u^\alpha_I) \in \EuScript{E}$. For the partial derivatives of $w^a$ which are defined as
$w^a_{x^k} =  \tilde{D}_{x^k}(w^a)$ we have the system of
{\it covering equations}
\[
w^a_{x^k} = T^a_k(x^i,u^\alpha_I,w^b).
\]
This over-determined system of {\sc pde}s is compatible whenever $(x^i,u^\alpha_I) \in \EuScript{E}$.

\vskip 5 pt

Dually the covering with extended total derivatives (\ref{extended_derivatives}) is defined by the differential
ideal generated by the {\it Wahlquist--Estabrook forms}, \cite[p.~81]{LNM810},
\[
\varpi^a = dw^a - \sum \limits_{k=1}^n T^a_k(x^i,u^\alpha_I,w^b)\,dx^k.
\]
This ideal is integrable on $\EuScript{E}$, that is,
\[
d \varpi^a \equiv \sum \limits_{b} \eta^{a}_{b} \wedge \varpi^b
\,\,\,\mathrm{mod}\,\, \langle\, \vartheta^\beta_I\,\rangle,
\]
where $\eta^a_b$ are some 1-forms on $\EuScript{E}\times \mathcal{W}$ and
$\vartheta^\beta_I = (du^\beta_I-\sum \limits_{k} u^\beta_{I+1_k}dx^k)\vert_{\EuScript{E}}$.

\subsection{Exotic cohomology}

Let $\mathfrak{g}$ be a Lie algebra over $\mathbb{R}$ and $\rho \colon \mathfrak{g} \rightarrow \mathrm{End}(V)$
be its representation. Let $C^k(\mathfrak{g}, V) =\mathrm{Hom}(\Lambda^k(\mathfrak{g}), V)$, $k \ge 1$,
be the space of all $k$--linear skew-symmetric mappings from $\mathfrak{g}$ to $V$. Then
the Chevalley--Eilenberg differential complex
\[
V=C^0(\mathfrak{g}, V) \stackrel{d}{\longrightarrow} C^1(\mathfrak{g}, V)
\stackrel{d}{\longrightarrow} \dots \stackrel{d}{\longrightarrow}
C^k(\mathfrak{g}, V) \stackrel{d}{\longrightarrow} C^{k+1}(\mathfrak{g}, V)
\stackrel{d}{\longrightarrow} \dots
\]
is generated by the differential defined as
\[
d \theta (X_1, ... , X_{k+1}) =
\sum\limits_{q=1}^{k+1}
(-1)^{q+1} \rho (X_q)\,(\theta (X_1, ... ,\hat{X}_q, ... ,  X_{k+1}))
\]
\begin{equation}
\quad
+\sum\limits_{1\le p < q \le k+1} (-1)^{p+q}
\theta ([X_p,X_q],X_1, ... ,\hat{X}_p, ... ,\hat{X}_q, ... ,  X_{k+1}).
\label{CE_differential}
\end{equation}
The cohomology groups of the complex $(C^{*}(\mathfrak{g}, V), d)$ are referred to as
the {\it cohomology groups of the Lie algebra} $\mathfrak{g}$ {\it with coefficents in the representation}
$\rho$. For the trivial representation $\rho_0 \colon \mathfrak{g} \rightarrow \mathbb{R}$,
$\rho_0 \colon X \mapsto 0$, the complex and its cohomology are denoted by $C^{*}(\mathfrak{g})$ and
$H^{*}(\mathfrak{g})$, respectively.

Consider a Lie algebra $\mathfrak{g}$ over $\mathbb{R}$ with non-trivial first cohomology group
$H^1(\mathfrak{g})$ and take a closed 1-form $\alpha$ on $\mathfrak{g}$ non-cohomologous to 0.
Then for any $\lambda \in \mathbb{R}$ define new differential
$d_{\lambda\alpha} \colon C^k(\mathfrak{g},\mathbb{R}) \rightarrow C^{k+1}(\mathfrak{g},\mathbb{R})$ by
the formula
\[
d_{\lambda \alpha} \theta = d \theta +\lambda \,\alpha \wedge \theta.
\]
From  $d\alpha = 0$ it follows that
\begin{equation}
d_{\lambda \alpha} ^2=0.
\label{d_deformed_2}
\end{equation}
The cohomology groups of the complex
\[
C^1(\mathfrak{g}, \mathbb{R})
\stackrel{d_{\lambda \alpha}}{\longrightarrow}
\dots
\stackrel{d_{\lambda \alpha}}{\longrightarrow}
C^k(\mathfrak{g}, \mathbb{R})
\stackrel{d_{\lambda \alpha}}{\longrightarrow}
C^{k+1}(\mathfrak{g}, \mathbb{R})
\stackrel{d_{\lambda \alpha}}{\longrightarrow} \dots
\]
are referred to as the {\it exotic} {\it cohomology groups} of $\mathfrak{g}$ and denoted by
$H^{*}_{\lambda\alpha}(\mathfrak{g})$, \cite{Novikov2002}.

\vskip 5 pt

\noindent
{\sc Remark 1}.
Cohomology $H^{*}_{\lambda\alpha}(\mathfrak{g})$ coincides with cohomology of $\mathfrak{g}$ with coefficients
in the one-dimensional representation $\rho_{\lambda\alpha} \colon \mathfrak{g} \rightarrow \mathbb{R}$,
$\rho_{\lambda\alpha} \colon X \mapsto \lambda\, \alpha(X)$. In particular,
when $\lambda=0$, cohomology $H^{*}_{\lambda\alpha}(\mathfrak{g})$ coincides with $H^{*}(\mathfrak{g})$.
\hfill$\diamond$

\vskip 5 pt

\noindent
{\sc Remark 2}.
In all the cases considered in this paper
$H^1(\mathfrak{g})=Z^1(\mathfrak{g})$ due to
$C^0(\mathfrak{g}) = \{0\}$ and $B^1(\mathfrak{g})=\{0\}$,
so a closed 1-form $\alpha$ can be identified with its cohomology class.
\hfill$\diamond$

\section{Reduced quasi-classical self-dual Yang--Mills equation and its integrable generalization}

\subsection{Symmetry algebra of rqsdYM}

The symmetry algebra $\mathfrak{s}_2$ of equation (\ref{FKh4}) admits generators
\begin{eqnarray*}
W_0(A) &=& -(A_z\,x+A_t\,y)\,u_x-A\,u_z+A_z\,u +\case{1}{2}\,A_{zz}\,x^2+A_{tz}\,x\,y,+\case{1}{2}\,A_{tt}\,y^2,
\\
W_1(A) &=& -A\,u_x+A_z\,x+A_t\,y,
\\
W_2(A) &=& A,
\\
X &=& -x\,u_x-y\,u_y + 2\,u,
\\
Y_0 &=& -u_t,
\\
Y_1 &=& - t\,u_t+\case{1}{2}\,(x\,u_x-y\,u_y)-u,
\\
Y_2 &=& -\case{1}{2}\,(t^2\,u_t+t\,x\,u_x-t\,y\,u_y-x\,y)-t\,u,
\\
Z_0 &=& - u_y,
\\
Z_1 &=& -t\,u_y-x,
\end{eqnarray*}
where $A = A(t,z)$ are arbitrary functions. These generators correspond to  point sym\-met\-ri\-es
$\hat{W}_i(A)$, $\hat{X}$,  $\hat{Y}_j$, $\hat{Z}_k$  defined in accordance with (\ref{hat_rule}). The
commutator table of $\mathfrak{s}_2$ is given by equa\-ti\-ons
\[[W_i(A), W_j(B)] =
\left\{
\begin{array}{lll}
W_{i+j}(A\,B_z-B\,A_z), &~~~& i+j \le 2
\\
0, && i+j>2,
\end{array}
\right.
\]
\[
[X, W_k(A)] = -k\,W_k(A),
\]
\[
[Y_0, W_k(A)] = W_k(A_t),
\]
\[
[Y_1, W_k(A)] = W_k(t\,A_t+\case{1}{2}\, k\,A),
\]
\[
[Y_2, W_k(A)] = \case{1}{2}\,W_k(t^2\,A_t+k\,t\,A),
\]
\[
[Z_0, W_k(A)] =
\left\{
\begin{array}{lll}
W_{k+1}(A_t), &~~~& k \le 1
\\
0, && k=2
\end{array}
\right.
\]
\[
[Z_1, W_k(A)] =
\left\{
\begin{array}{lll}
W_{k+1}(t\,A_t+k\,A), &~~~& k \le 1
\\
0, && k=2
\end{array}
\right.
\]
\[
\begin{array}{lclcl}
[X, Y_m]  = 0,       & ~~~ &  [X, Z_m] = - Z_m,  & ~~~ & [Z_0, Z_1]=0,
\\   {}
[Y_0, Y_1] = Y_0,  &         &  [Y_0, Y_2]= Y_1,   &         & [Y_1, Y_2] = Y_2,
\\ {}
[Y_0, Z_0] =0, && [Y_1, Z_0] = -\frac{1}{2}\,Z_0, && [Y_2, Z_0] = -\frac{1}{2}\,Z_1,
\\ {}
[Y_0, Z_1] =Z_0, && [Y_1, Z_1] = \frac{1}{2}\,Z_1, && [Y_2, Z_1] =0.
\end{array}
\]
From this table it follows that the symmetry algebra of equation (\ref{FKh4}) is  the semi-direct sum
$\mathfrak{s}_2 = \mathfrak{s}_{\diamond} \ltimes \mathfrak{s}_{2,\infty}$
of the finite-dimensional Lie algebra $\mathfrak{s}_{\diamond}$ generated by $X$, $Y_i$, $Z_j$, and the
infinite-dimensional ideal  $\mathfrak{s}_{2,\infty}$ generated by $W_0(A)$, $W_1(A)$, $W_2(A)$.
We  have
$\mathfrak{s}_{\diamond} = \mathfrak{a} \ltimes (\mathfrak{sl}_2(\mathbb{R}) \ltimes \mathfrak{b})$,
where $\mathfrak{a} =\langle X\rangle$,
$\mathfrak{sl}_2(\mathbb{R})  = \langle Y_0, Y_1, Y_2\rangle$, and
$\mathfrak{b} = \langle Z_0, Z_1\rangle$ is a two-dimensional  Abelian Lie algebra,
while $\mathfrak{s}_{2,\infty}$ is isomorphic to the  tensor product
$\mathfrak{q} \otimes \mathbb{R}_2[h]$
of the Lie algebra of $\mathfrak{q}$ of the vector fields of the form
$A(t,z)\,\partial_z$ on $\mathbb{R}^2$ and the associative commutative unital algebra of truncated
polynomials $\mathbb{R}_2[h] = \mathbb{R}[h]/\langle h^3 \rangle $
of order less than 3 in the (formal) variable $h$.

\subsection{Maurer--Cartan forms and non-triviality of the second exotic cohomology group of $\mathfrak{s}_2$}
\label{MCfs_subsection}

Consider the Maurer--Cartan forms  $\alpha$, $\beta_i$, $i \in \{0, 1, 2\}$ $\gamma_l$, $l \in \{0, 1\}$, $\theta_{k,m,n}$,
$k\in \{0, 1, 2\}$, $m, n \in \mathbb{N} \cup \{0\}$, of the Lie algebra $\mathfrak{s}_2$, that are dual to  the basis
$X$, $Y_i$, $Z_l$, $W_k(t^mz^n)$  of $\mathfrak{s}_2$; in other words, take 1-forms such that there hold
$\alpha(\hat{X}) =1$,
$\beta_{i^{\prime}}(\hat{Y}_{i^{\prime\prime}}) = \delta_{i^{\prime}i^{\prime\prime}}$,
$\gamma_{l^{\prime}}(\hat{Z}_{l^{\prime\prime}}) = \delta_{l^{\prime}l^{\prime\prime}}$,
$\theta_{k^{\prime},m^{\prime},n^{\prime}}(\hat{W}_{k^{\prime\prime}}(t^{m^{\prime\prime}}z^{n^{\prime\prime}}))
= \delta_{k^{\prime}k^{\prime\prime}} \,\delta_{m^{\prime}m^{\prime\prime}} \,\delta_{n^{\prime}n^{\prime\prime}}$,
while all the other values of these 1-forms on the elements of the basis are equal to zero. Denote
\[
B = \beta_0 + h_1\,\beta_1+\case{1}{2}\,h_1^2 \beta_2,
\qquad
\Gamma = \gamma_0 + h_1\,\gamma_1,
\]
and consider the formal series of 1-forms
\[
\Theta_k =
\sum \limits_{m=0}^{\infty}\sum \limits_{n=0}^{\infty} \frac{h_1^m}{m!}\frac{h_2^n}{n!}\, \theta_{k,m,n},
\]
where $h_1$ and $h_2$ are formal parameters such that $dh_1 = dh_2 =0$. Then (\ref{CE_differential})
and the commutator relations of $\mathfrak{s}_2$ yield  Cartan's structure equations
\begin{eqnarray}
\fl
d\alpha &=& 0,
\label{da_eq}
\\
\fl
dB &=& \nabla_1(B) \wedge B,
\label{dB_eq}
\\
\fl
d\Gamma &=& \alpha \wedge \Gamma + \nabla_1(\Gamma) \wedge B +\case{1}{2}\,\nabla_1(B) \wedge \Gamma,
\label{dG_eq}
\\
\fl
d\Theta_k &=& k\,\left(\alpha - \case{1}{2}\,\nabla_1(B)\right)\wedge \Theta_k
+\sum\limits_{m=0}^k \nabla_2(\Theta_{k-m}) \wedge \Theta_m
+\nabla_1(\Theta_k) \wedge B +\nabla_1(\Theta_{k-1}) \wedge \Gamma
\nonumber
\\
\fl
&&
- (k-1)\,\nabla_1(\Gamma) \wedge \Theta_{k-1},
\label{dT_k_eq}
\end{eqnarray}
where $k \in \{0, 1, 2\}$, $\Theta_{-1} = 0$,  and $\nabla_{l}=\frac{\partial}{\partial h_l}$.

\vskip 7 pt
From the structure equations (\ref{da_eq}), (\ref{dB_eq}), (\ref{dG_eq}) of the Lie algebra
$\mathfrak{s}_\diamond$ we have the following theorem, which can be proved by direct computations:
\vskip 7 pt
\noindent
{\sc Theorem 1}. $H^1(\mathfrak{s}_2) = \mathbb{R} [\alpha]$,
\[
H_{\lambda\,\alpha}^2(\mathfrak{s}_{\diamond}) =
\left\{
\begin{array}{lcl}
\mathbb{R}\,[\gamma_0 \wedge \gamma_1], &~~~& \lambda=-2,
\\
\{0\}, && \lambda \neq -2.
\end{array}
\right.
\]

\vskip 7 pt

\noindent
{\sc Corollary}.
{\it
Equation
\begin{equation}
d\sigma = 2\,\alpha \wedge \sigma + \gamma_0 \wedge \gamma_1
\label{main_extension}
\end{equation}
with unknown 1-form $\sigma$ is compatible with the structure equations
(\ref{da_eq}), (\ref{dB_eq}), (\ref{dG_eq}), (\ref{dT_k_eq})  of the Lie algebra $\mathfrak{s}_2$.
}
\vskip 7 pt

We can find all the Maurer--Cartan forms $\alpha$, $\beta_i$, $\gamma_l$, $\theta_{k,m,n}$, and the additional
form $\sigma$ by integration of the structure equations (\ref{da_eq}), (\ref{dB_eq}), (\ref{dG_eq}),
(\ref{dT_k_eq}), (\ref{main_extension}). For the purposes of this paper we need the forms  $\alpha$, $\beta_i$,
$\gamma_l$, $\theta_{k,0,0}$, $\sigma$ only. We have consequently
\[\fl
\alpha =\frac{d a_0}{a_0},
\qquad
\beta_0 = a_1^2\,dt,
\qquad
\beta_1 = 2\,\frac{d a_1}{a_1}+a_2\,dt,
\qquad
\beta_2 = \frac{1}{a_1^2}\,\left(d a_2 +\frac{a_2^2}{2}\,dt\right),
\]
\[\fl
\gamma_0 = a_0\,a_1\,\left(dy + a_3 \,dt\right),
\qquad
\gamma_1 = \frac{a_0}{a_1}\,\left(da_3 + \case{1}{2}\,a_2 \,(dy + a_3\,dt)\right),
\]
\[\fl
\sigma = a_0^2\,\left(dv-a_3\,dy-\case{1}{2}\,a_3^2\,dt\right),
\qquad
\theta_{0,0,0}= b_0 \,dz+b_1\,dt,
\]
\[\fl
\theta_{1,0,0} = \frac{a_0 b_0}{a_1}\,\left(dx +\frac{b_1}{b_0}\,dy + b_2\,dz+b_3\,dt\right)
\]
\begin{equation}
\fl
\theta_{2,0,0} = \frac{a_0^2\,b_0}{a_1^2}\,\left(du+(b_2-a_3)\,dx+\left(b_3-\frac{a_3 b_1}{a_0}\right)\,dy+b_4\,dz+b_5\,dt\right),
\label{t200_3_2}
\end{equation}
where $a_0$, ..., $a_3$, $b_0$, ... , $b_5$, $t$, $x$, $y$, $z$, $u$, $v$ are parameters (``constants of integration'')
such that\footnote[7]{we put $\alpha = da_0/a_0$ instead of the natural choice $\alpha = da_0$ and
$\beta_0 = a_1^2\,dt$, instead of $\beta_0 = a_1\,dt$ to simplify the further computations}
$a_0 \neq0$,
$a_1 \neq 0$, and $b_0 \neq 0$.
Then we  rename parameters in the form $\theta_{2,0,0}$  to make it to be a contact form
\[
\theta_{2,0,0} = \frac{a_0^2\,b_0}{a_1^2}\,(du - u_t\,dt-u_x\,dx-u_y\,dy-u_z\,dz),
\]
of the  order 0 on the bundle
$J^1(\pi)$,
$\pi \colon \mathbb{R}^4 \rightarrow \mathbb{R}^3$,
$\pi \colon (t,x,y,z,u) \mapsto (t,x,y,z)$, that is, we put
$b_2 = a_3 - u_x$, $b_3 = -u_y+a_3 b_1a_0^{-1}$, $b_4 = -u_z$, $b_5 =- u_t$.
Then we have
\[\fl
\sigma-\theta_{1,0,0} =
a_0^2\,\left(dv
-\frac{a_3\,(a_0^2 a_1 a_3+2\,b_0 b_1)}{2\, a_0^2a_1}\,dt
-\frac{b_0}{a_0 a_1}\,dx
-\frac{a_1 a_3+b_1}{a_0 a_1}\,dy
+\frac{b_0\,(u_x-a_3)}{a_0 a_1}\,dz\right).
\]
We introduce new parameters $v_x$, $v_y$, $w$ such that
\[
b_0 = a_0\,a_1\,v_x,
\quad
b_1 = a_1\,(a_0\,v_y-w),
\quad
a_3=w.
\]
This gives
\begin{equation}
\fl
\sigma-\theta_{1,0,0} = a_0^2\,\left(
dv
-\left(w\,v_y-u_y\,v_x-\case{1}{2}\,w^2\right)\,dt
- v_x\,dx-v_y\,dy -(w-u_x)\,v_x\,dz
\right).
\label{WE_form_of extended_FKh4}
\end{equation}

\subsection{Lax representation of a generalization of rqsdYM}

The 1-form (\ref{WE_form_of extended_FKh4}) is equal to zero whenever the following over-determined system holds:
\begin{equation}
\left\{
\begin{array}{lcl}
v_t &=& w\,v_y-u_y\,v_x-\case{1}{2}\,w^2,
\\
v_z &=& (w-u_x)\,v_x.
\end{array}
\right.
\label{extension_1_of_FKH4_covering}
\end{equation}
In the case of $w=\lambda = \mathrm{const}$ system (\ref{extension_1_of_FKH4_covering}) after the change of variable
$v = q-\case{1}{2}\,\lambda^2\,t$ coincides with system (\ref{FKH4_covering_lambda}). In the general case the compatibility condition
$(v_t)_z = (v_z)_t$ of system (\ref{extension_1_of_FKH4_covering}) gives an over-determined system
\begin{equation}
\left\{
\begin{array}{lcl}
w_t &=& w\,w_y-u_y\,w_x-(u_{yz}-u_{tx} - u_y\,u_{xx}+u_x\,u_{xy}),
\\
w_z &=& (w-u_x)\,w_x.
\end{array}
\right.
\label{extension_2_of_FKH4_covering}
\end{equation}
In its turn this system is compatible whenever
\[
\left\{
\begin{array}{lcl}
(u_{yz}-u_{tx} - u_y\,u_{xx}+u_x\,u_{xy})_x&=&0,
\\
(u_{yz}-u_{tx} - u_y\,u_{xx}+u_x\,u_{xy})_z&=&0.
\end{array}
\right.
\]
In other words, system (\ref{extension_2_of_FKH4_covering}) defines a covering over
the two-component generalization
\begin{equation}
\left\{
\begin{array}{lcl}
u_{yz} &=& u_{tx} + u_y\,u_{xx}-u_x\,u_{xy} + s,
\\
s_x &=& 0,
\\
s_z &=&0.
\end{array}
\right.
\label{FKh4_extended}
\end{equation}
of equation (\ref{FKh4}) and can be written in the form
\[
\left\{
\begin{array}{lcl}
w_t &=& w\,w_y-u_y\,w_x-s,
\\
w_z &=& (w-u_x)\,w_x.
\end{array}
\right.
\]
In the particular case $s=0$ we have the  covering over (\ref{FKh4}) defined by system
\begin{equation}
\left\{
\begin{array}{lcl}
w_t &=& w\,w_y-u_y\,w_x,
\\
w_z &=& (w-u_x)\,w_x.
\end{array}
\right.
\label{FKh4_eversed_covering}
\end{equation}
This system is related with system (\ref{FKH4_covering_lambda}) by the following transformation, cf.
\cite{PavlovChangChen2009}: suppose that a solution $q$ to (\ref{FKH4_covering_lambda})  is defined
implicitly as  $W(t,x,y,z,q(t,x,y,z))=\lambda =\mathrm{const}$, then $W$ is a solution to
(\ref{FKh4_eversed_covering}).

\vskip 5 pt
\noindent
{\sc Remark 3}.
Function $s$ can not be excluded from system (\ref{FKh4_extended}) by a contact trans\-for\-ma\-ti\-on,
therefore equations (\ref{FKh4}) and (\ref{FKh4_extended})  are not equivalent.
\hfill $\diamond$

\section{Integrable hierarchy associated to rqsdYM}

The Lie algebra $\mathfrak{s}_2$ admits a sequence of natural extensions
$\mathfrak{s}_m = \mathfrak{s}_\diamond \ltimes \mathfrak{s}_{m, \infty}$, $m\ge 3$,
where
\[
\mathfrak{s}_{m,\infty} =\{A(t,z)\,\partial_z\}  \otimes \mathbb{R}[h]/\langle h^{m+1} \rangle,
\]
with the structure equations of the form (\ref{da_eq}),  (\ref{dB_eq}), (\ref{dG_eq}), (\ref{dT_k_eq}) such
that $k \in \{0, \dots, m\}$ in  (\ref{dT_k_eq}). Since the finite-dimensional part in all the algebras
$\mathfrak{s}_m$ is the same, every Lie algebra $\mathfrak{s}_m$ satisfies Theorem 1 and  its Corollary.
Therefore we can find forms $\theta_{k,0,0}$, $k \le m$ by integration of the structure equations in the same
way as in Subsection \ref{MCfs_subsection}. Then we consider $\theta_{m,0,0}$ as a multiple of the contact form
$du -u_{x_1} dx_1 - \dots - u_{x_{m+2}} dx_{m+2}$ of order 0 on the jet bundle $J^1(\pi)$ for
$\pi \colon \mathbb{R}^{m+2} \times \mathbb{R} \rightarrow \mathbb{R}^{m+2}$,
$\pi \colon (x_1, \dots, x_{m+2}, u) \mapsto (x_1, \dots, x_{m+2})$,
take the 1-form $\sigma-\theta_{m-1,0,0}$ as the Wahlquist-Estabrook form of a covering and write out the
compatibility conditions explicitly.  In this Section we consider the cases $m=3$, $m=4$, and $m=5$.
Below we alter notation as follows: $t\mapsto x_1$, $x \mapsto x_2$, $y \mapsto x_3$, $z \mapsto x_4$.

\subsection{Case $k \in \{0, \dots, 3\}$.}

While the 1-forms $\alpha$, $\beta_i$, $\gamma_l$, $\sigma$, $\theta_{0,0,0}$, $\theta_{1,0,0}$ are the same
as in Subsection \ref{MCfs_subsection}, instead of (\ref{t200_3_2}) we have now
\[
\theta_{2,0,0} = \frac{a_0^2b_0}{a_1^2}\,
\left(
dx_5 + (b_2-a_3)\,dx_2 + \left(b_3-\frac{a_3 b_1}{b_0}\right)\,dx_3 + b_4 \, dx_4 +b_5 dx_1
\right).
\]
Then we put
\[
b_2 = 2\,a_3 - u_{x_5},
\quad
b_4 = a_3^2-a_3\,u_{x_5}-u_{x_2},
\quad
b_5 = a_3\,b_3 -\frac{a_3^3b_1}{b_0}-u_{x_3}
\]
and obtain
\[
\theta_{3,0,0} = \frac{a_0^3b_0}{a_1^3}\left(
du-\sum \limits_{i=1}^5 u_{x_i}\,dx_i
\right).
\]
Further we rename
$b_0=a_1^2\,v_{x_5}$,
$a_3=v_{x_2}v_{x_5}^{-1}+u_{x_5}$,
$b_3=(a_1^2\,(v_{x_5}\,(v_{x_3}-u_{x_5})-v_{x_1})+b_1\,(v_{x_2}+u_{x_5}\,v_{x_5}))\,a_1^{-2}v_{x_5}^{-2}$.
This yields
\[
\fl
\sigma - \theta_{2,0,0} =
a_0^2\,\left(
dv- v_{x_2}\,dx_2-v_{x_3}\,d{x_3}-v_{x_5}\,dx_5
- \frac{v_{x_2}^2+u_{x_5}\,v_{x_2}\,v_{x_5}-u_{x_2}\,v_{x_5}^2}{v_{x_5}}\,dx_4
\right.
\]
\[
\fl\qquad\qquad\qquad\qquad
\left.
-\left(u_{x_5}\,v_{x_3} +\frac{(v_{x_3}-u_{x_5})\,v_{x_2}}{v_{x_5}}-u_{x_3}\,v_{x_5}
-\frac{u_{x_5}^2\,v_{x_5}^2-v_{x_2}^2}{2\,v_{x_5}^2}\right)\,dx_1
\right).
\]
This 1-form is equal to zero whenever there holds the over-determined system
\[
\left\{
\begin{array}{lcl}
v_{x_1} &=& \displaystyle{u_{x_5}\,v_{x_3} +\frac{(v_{x_3}-u_{x_5})\,v_{x_2}}{v_{x_5}}-u_{x_3}\,v_{x_5}
-\frac{u_{x_5}^2\,v_{x_5}^2-v_{x_2}^2}{2\,v_{x_5}^2}},
\\
v_{x_4} &=& \displaystyle{\frac{v_{x_2}^2+u_{x_5}\,v_{x_2}\,v_{x_5}-u_{x_2}\,v_{x_5}^2}{v_{x_5}}}.
\end{array}
\right.
\]
 The compatibility condition $(v_{x_1})_{x_4} = (v_{x_4})_{x_1}$ of this system gives three  equations
for the function $u$:
\begin{eqnarray}
u_{x_4x_5} &=& u_{x_2x_2}-u_{x_2}\,u_{x_5x_5}+u_{x_5}\,u_{x_2x_5},
\label{Pavlov_changed}
\\
u_{x_1x_5} &=& u_{x_2x_3}-u_{x_3}\,u_{x_5x_5}+u_{x_5}\,u_{x_3x_5},
\label{FKh4_changed}
\\
u_{x_3x_4} &=& u_{x_1x_2}+u_{x_3}\,u_{x_2x_5} - u_{x_2}\,u_{x_3x_5}.
\label{MASh5_changed}
\end{eqnarray}
This system is compatible. Equations (\ref{FKh4_changed}) and  (\ref{MASh5_changed}) differ from equations
(\ref{FKh4}) and  (\ref{MASh5}) only by notation,  while equation (\ref{Pavlov_changed}) was introduced in
\cite{Kuzmina1967} and is known to have a covering with non-removable parameter,
see \cite{Mikhalev1992,Pavlov2003,Dunajski2004}.

\subsection{Case $k \in \{0, \dots, 4\}$.}

In this case we get
\vskip 5 pt
\[
\fl
\theta_{3,0,0} =
\frac{a_0^3b_0}{a_1^3}\,\left(
dx_6 +(b_3-2\,a_3)\,dx_5 +(b_4-a_3b_2+a_3^2)\,dx_2
+(b_5-a_3 b_3+b_1 a_3^2)\,dx_3
\right.
\]
\[
\fl\qquad\qquad
\left.
+b_6\,dx_4+b_7\,dx_1
\right),
\]
then substituting for  $b_2=3\,a_3-u_{x_6}$,
$b_4=-u_{x_5}-2\,a_3\,u_{x_6}+3\,a_3^2$,
$b_7=a_3^3-u_{x_2}-a_3^2\,u_{x_6}-a_3\,u_{x_5}$ into $\theta_{4,0,0}$
yields
$\theta_{4,0,0} = a_0^4b_0 a_1^{-4}\left(du-u_{x_1}\,dx_1 - \dots - u_{x_6}\, dx_6\right)$.
Further we introduce new parameters  $v_{x_3}$, $v_{x_5}$, $v_{x_6}$ such that
$b_0=a_1^3a_0^{-1}\,v_{x_6}$,
$a_3=v_{x_5}v_{x_6}^{-1}+u_{x_6}$,
$b_5 =(v_{x_6}\,(v_{x_6}\,u_{x_6}+v_{x_5})\,b_3 -(v_{x_6}\,u_{x_6}+v_{x_5})^2\,b_1-v_{x_5}$
$-  v_{x_6}\,(u_{x_6}-v_{x_3}))\,v_{x_6}^{-2}$. This gives
\[
\fl
\sigma-\theta_{3,0,0}=a_0^3\,\left(
dv
- v_{x_3} dx_3
- v_{x_5} dx_5
-v_{x_6} dx_6
- \frac{v_{x_5}^2+u_{x_6} v_{x_5} v_{x_6}-u_{x_5} v_{x_6}^2}{v_{x_6}^2}\,dx_2
\right.
\]
\[
\fl
\qquad\qquad
- \left(
u_{x_6} v_{x_5}-\frac{v_{x_5}\,(v_{x_5}+2\,(u_{x_6}-v_{x_3})\,v_{x_6})}{2\,v_{x_6}^2}
-u_{x_3} v_{x_6}-\frac{u_{x_6}^2}{2}
\right)\,dx_1
\]
\[
\fl\qquad\qquad
\left.
 - \left(\frac{v_{x_5}^2\,(v_{x_5}+2\,u_{x_6} v_{x_6})}{v_{x_6}^2}
-(u_{x_5} u_{x_6} +u_{x_2})\,v_{x_6}-(u_{x_5}-u_{x_6}^2)\,v_{x_5}
\right)\,dx_4
\right).
\]
This 1-form defines an over-determined system
\begin{equation}
\left\{
\begin{array}{lcl}
v_{x_1} &=&
\displaystyle{
u_{x_6} v_{x_5}-\frac{v_{x_5}\,(v_{x_5}+2\,(u_{x_6}-v_{x_3})\,v_{x_6})}{2\,v_{x_6}^2}
-u_{x_3} v_{x_6}-\frac{u_{x_6}^2}{2},
}
\\
v_{x_2} &=&
\displaystyle{
\frac{v_{x_5}^2+u_{x_6} v_{x_5} v_{x_6}-u_{x_5} v_{x_6}^2}{v_{x_6}^2},
}
\\
v_{x_4} &=&
\displaystyle{
\frac{v_{x_5}^2\,(v_{x_5}+2\,u_{x_6} v_{x_6})}{v_{x_6}^2}
-(u_{x_5} u_{x_6} +u_{x_2})\,v_{x_6}-(u_{x_5}-u_{x_6}^2)\,v_{x_5}.
}
\end{array}
\right.
\label{k=4_covering}
\end{equation}
System  (\ref{k=4_covering}) is compatible by virtue of the following system of the second order equations for function $u$:
\begin{eqnarray}
u_{x_5x_5} &=& u_{x_2x_6} +u_{x_5} \,u_{x_6x_6} - u_{x_6}\,u_{x_5x_6},  
\label{Pavlov_changed_2}
\\
u_{x_4x_6} &=& u_{x_2x_5} +u_{x_6} \,u_{x_2x_6} - u_{x_2}\,u_{x_6x_6},   
\label{FKh4_changed_2}
\\
u_{x_3x_5} &=&u_{x_1x_6} +u_{x_3} \,u_{x_6x_6} - u_{x_6}\,u_{x_3x_6}, 
\label{FKh4_changed_3}
\\
u_{x_2x_3}&=&u_{x_1x_5} +u_{x_3} \,u_{x_5x_6} - u_{x_5}\,u_{x_3x_6},  
\label{MASh5_changed_2}
\\
u_{x_3x_4}&=&u_{x_1x_2} +u_{x_3} \,u_{x_2x_6} - u_{x_2}\,u_{x_3x_6}, 
\label{MASh5_changed_3}
\\
u_{x_4x_5} &=& u_{x_2x_2} +u_{x_5} \,u_{x_2x_6} - u_{x_2}\,u_{x_5x_6}.   
\label{Pavlov_Stoilov}
\end{eqnarray}
The last system is compatible. Equations (\ref{Pavlov_changed_2}), (\ref{FKh4_changed_2}),
(\ref{FKh4_changed_3}), (\ref{MASh5_changed_2}), (\ref{MASh5_changed_3})   differ from equations
(\ref{Pavlov_changed}), (\ref{FKh4}),  (\ref{FKh4}), (\ref{MASh5}), (\ref{MASh5}), respectively, by notation.
Equation (\ref{Pavlov_Stoilov}) was introduced in \cite{PavlovStoilov2017}, where a covering with a
non-removable parameter for this equation was presented.

\subsection{Case $k \in \{0, \dots, 5\}$.}

In this case we have
\[
\fl
\theta_{4,0,0} =
\frac{a_0^4b_0}{a_1^4}\,\left(
dx_7
+b_{10}\,dx_1
+b_{9}\,dx_4
+(b_2-3\,a_3)\,dx_6
+(b_4-2\,a_3\,b_2+3\,a_3^2)\,dx_5
\right.
\]
\[
\fl
\qquad\qquad
\left.
+(b_7-a_3\,b_4+a_3^2\,b_2-a_3^3)\,dx_2
+(b_8-a_3\,b_5+a_3^2\,b_3-a_3^3\,b_1)\,dx_3
\right).
\]
Then after altering  notation
$b_4=6\,a_3^2-3\,a_3\,u_{x_7}-u_{x_6}$,
$b_7=4\,a_3^3-3\,a_3^2\,u_{x_7}-2\,a_3\,u_{x_6}-u_{x_5}$,
$b_9=a_3^4-a_3^3\,u_{x_7}-a_3^2\,u_{x_6}-a_3\,u_{x_5}-u_{x_2}$,
$b_{10}=-a_3^4\,b_1+a_3^3\,b_3-a_3^2\,b_5+a_3\,b_8-u_{x_3}$
we obtain
$\theta_{5,0,0} = a_0^5b_0 a_1^{-5}\left(du-u_{x_1}\,dx_1 - \dots - u_{x_7}\, dx_7\right)$.
Further we rename
$b_0=v_{x_7}\,a_1^4 a_0^{-2}$,
$b_2=4\,a_3-u_{x_7}$,
$a_3=v_{x_6}v_{x_7}^{-1}+u_{x_7}$,
$b_8 = b_5\,u_{x_7}-b_3\,u_{x_7}^2+b_1\,u_{x_7}^3
+(b_5\,v_{x_6}-2\,b_3\,v_{x_6}\,u_{x_7}+v_{x_3}+3\,b_1\,v_{x_6}\,u_{x_7}^2-u_{x_7})\,v_{x_7}^{-1}
+(3\,b_1\,v_{x_6}^2\,u_{x_7}-v_{x_6}-b_3\,v_{x_6}^2)\,v_{x_7}^{-2}+b_1\,v_{x_6}^3\,v_{x_7}^{-3}$.
This gives
\[\fl
\sigma -\theta_{4,0,0} = a_0^2\,(
dv - v_{x_3} dx_3 - v_{x_6} dx_6 - v_{x_7} dx_7
-(u_{x_7}\,v_{x_6}-u_{x_6}\,v_{x_7}+v_{x_6}^2\,v_{x_7}^{-1})\,dx_5
\]
\[
-(u_{x_3}\,v_{x_7}-u_{x_7}\,v_{x_3}+v_{x_6}\,(u_{x_7}-v_{x_3})\,v_{x_7}^{-1}
+\case{1}{2}\,(u_{x_7}^2+v_{x_6}^2\,v_{x_7}^{-2}))\, dx_1
\]
\[
-((u_{x_7}^2-u_{x_6})\,v_{x_6}+2\,v_{x_6}^2\,u_{x_7}\,v_{x_7}^{-1}
+v_{x_6}^3\,v_{x_7}^{-2}-(u_{x_7}\,u_{x_6}+u_{x_5})\,v_{x_7})\,dx_2
\]
\[
-((u_{x_7}^3-2\,u_{x_7}\,u_{x_6}-u_{x_5})\,v_{x_6}
+(3\,u_{x_7}^2-u_{x_6})\,v_{x_6}^2\,v_{x_7}^{-1}
+3\,u_{x_7}v_{x_6}^3\,v_{x_7}^{-2}
\]
\[
\qquad
+v_{x_6}^4\,v_{x_7}^{-3}
-(u_{x_7}^2\,u_{x_6}+u_{x_7}\,u_{x_5}+u_{x_2})\,v_{x_7}
)\,dx_4
).
\]
This 1-form produces the over-determined system
\[\fl
\left\{
\begin{array}{lcl}
v_{x_1} &=& u_{x_3}\,v_{x_7}-u_{x_7}\,v_{x_3}+v_{x_6}\,(u_{x_7}-v_{x_3})\,v_{x_7}^{-1}
+\case{1}{2}\,(u_{x_7}^2+v_{x_6}^2\,v_{x_7}^{-2}),
\\
v_{x_2} &=&
(u_{x_7}^2-u_{x_6})\,v_{x_6}+2\,v_{x_6}^2\,u_{x_7}\,v_{x_7}^{-1}
+v_{x_6}^3\,v_{x_7}^{-2}-(u_{x_7}\,u_{x_6}+u_{x_5})\,v_{x_7},
\\
v_{x_4} &=&(u_{x_7}^3-2\,u_{x_7}\,u_{x_6}-u_{x_5})\,v_{x_6}
+(3\,u_{x_7}^2-u_{x_6})\,v_{x_6}^2\,v_{x_7}^{-1}
+3\,u_{x_7}v_{x_6}^3\,v_{x_7}^{-2}+v_{x_6}^4\,v_{x_7}^{-3}
\\
&&
-(u_{x_7}^2\,u_{x_6}+u_{x_7}\,u_{x_5}+u_{x_2})\,v_{x_7},
\\
v_{x_5} &=& u_{x_7}\,v_{x_6}-u_{x_6}\,v_{x_7}+v_{x_6}^2\,v_{x_7}^{-1}.
\end{array}
\right.
\]
The compatibility conditions of this system yield the following equations of the second order for function $u$:
\begin{eqnarray}
\fl
u_{x_6x_6} &=& u_{x_5x_7}-u_{x_7}\,u_{x_6x_7}+u_{x_6}\,u_{x_7x_7},  
\label{Pavlov_changed_3}
\\
\fl
u_{x_1x_7} &=& u_{x_3x_6}+u_{x_7}\,u_{x_3x_7}-u_{x_3}\,u_{x_7x_7},  
\label{FKh4_changed_4}
\\
\fl
u_{x_2x_7} &=& u_{x_5x_6}+ u_{x_7}\,u_{x_5x_7}-u_{x_5}\,u_{x_7x_7},  
\label{FKh4_changed_5}
\\
\fl
u_{x_1x_5} &=&u_{x_2x_3} -u_{x_3}\,u_{x_5x_7}+u_{x_5}\,u_{x_3x_7},  
\label{MASh5_changed_5}
\\
\fl
u_{x_1x_6} &=& u_{x_3x_5}+u_{x_6}\,u_{x_3x_7}-u_{x_3}\,u_{x_6x_7},  
\label{MASh5_changed_6}
\\
\fl
u_{x_2x_6} &=&u_{x_4x_7}
-u_{x_7}^2\,u_{x_5x_7}
-u_{x_7}\,u_{x_5x_6}
+ (u_{x_5}\,u_{x_7}+u_{x_2})\,u_{x_7x_7},    
\label{New_5D_1}
\\
\fl
u_{x_4x_5} &=& u_{x_2x_2}
+u_{x_5}\,u_{x_5x_6}
+(u_{x_5}\,u_{x_7}-u_{x_2})\,u_{x_5x_7}
-u_{x_5}^2\,u_{x_7x_7}   
\label{New_5D_2}
\\
\fl
u_{x_4x_6} &=& u_{x_2x_5}
+u_{x_6}\,u_{x_5x_6}+u_{x_6}\,u_{x_7}\,u_{x_5x_7}-u_{x_2}\,u_{x_6x_7}
-u_{x_5}\,u_{x_6}\,u_{x_7x_7},    
\label{New_5D_3}
\\
\fl
u_{x_5x_5} &=&
u_{x_4x_7}
-u_{x_7}\,u_{x_5x_6}
-(u_{x_7}^2+u_{x_6})\,u_{x_5x_7}
+u_{x_5}\,u_{x_6x_7}
+(u_{x_5}\,u_{x_7}+u_{x_2})\,u_{x_7x_7},    
\label{New_5D_4}
\\
\fl
u_{x_3x_4} &=&
u_{x_1x_2}-u_{x_2}\,u_{x_3x_7}+u_{x_3}\,u_{x_5x_6}
+u_{x_3}\,u_{x_7}\,u_{x_5x_7}-u_{x_3}\,u_{x_5}\,u_{x_7x_7},   
\label{New_7D}
\end{eqnarray}
In its turn this system is compatible.  Equations (\ref{Pavlov_changed_3}), (\ref{FKh4_changed_4}), (\ref{FKh4_changed_5}), (\ref{MASh5_changed_5}), (\ref{MASh5_changed_6}) differ from equations (\ref{Pavlov_changed}),  (\ref{FKh4}), (\ref{FKh4}), (\ref{MASh5}), (\ref{MASh5}), respectively, by notation. We have not found equations (\ref{New_5D_1}), (\ref{New_5D_2}), (\ref{New_5D_3}), (\ref{New_5D_4}) with five independent variables and equation (\ref{New_7D}) with seven independent variables in the literature.

\section*{Acknowledgments}

This work was partially supported by the Faculty of Applied Mathematics of AGH UST statutory tasks within
subsidy of Ministry of Science and Higher Education.

I am very grateful to I.S. Krasil${}^{\prime}$shchik for useful discussions.
I thanks L.V. Bog\-da\-nov  for important remarks.

\end{document}